\documentclass[a4paper, reqno,10pt]{amsart}
\usepackage{amssymb}
\usepackage{graphicx}
\usepackage{verbatim}

\newcommand{\lap}{\bigtriangleup}
\newcommand{\grad}{\bigtriangledown}

\def\CC{{\rm\kern.24em\vrule width.02em height1.4ex
depth-.05ex\kern-.26em C}}
\def\RR{{\rm\kern.12em\vrule width.01em height1.4ex
depth-.05ex\kern-.26em R}}

\newcommand{\bdy}{\partial}
\newcommand{\n}{\noindent}

\newbox\qedbox
{\hfill\penalty10000\copy\qedbox\par\medskip}
      {\smallskip\noindent{\bf Proof of #1.}\hskip \labelsep}%
                        {\nobreak\hfill\hfill\nobreak\copy\qedbox\par\medskip}

\newcommand{\bbR}{\mathbb{R}}

\newcommand{\cH}{\mathcal{H}}

\newcommand{\cT}{\mathcal{T}}
\newcommand{\cR}{\mathcal{R}}

\newcommand{\fX}{\mathfrak{X}}

\newcommand{\fU}{\mathfrak{U}}

\newcommand{\bx}{{\mathbf x}}
\newcommand{\be}{{\mathbf e}}
\newcommand{\by}{{\mathbf y}}

\newcommand{\ba}{{\mathbf a}}
\newcommand{\bb}{{\mathbf b}}

\newcommand{\bw}{{\mathbf w}}

\newcommand{\bY}{{\mathbf Y}}
\newcommand{\bX}{{\mathbf X}}

\newcommand{\bW}{{\mathbf W}}

\newcommand{\bE}{{\mathbf E}}

\newcommand{\bF}{{\mathbf F}}
\newcommand{\bR}{{\mathbf R}}

\newcommand{\ud}{\mathrm d}

\newcommand{\but }{\! \bullet \!}

\DeclareMathOperator{\dist}{dist}

\begin{document}
\title[Free Energy]{Protein Folding: The Gibbs Free Energy}
\author{Yi Fang}
\address{Department of Mathematics \\
         Nanchang University \\
         999 Xuefu Road, Honggutan New District \\
         Nanchang, China, 330031 \\
         yi.fang3@gmail.com}
\date{March 21, 2012.} 
\maketitle

\setlength{\baselineskip}{1.2\baselineskip}
\begin{abstract}
   The fundamental law for protein folding is the {\bf Thermodynamic 
   Principle}:  the amino acid sequence of a protein determines its native 
   structure and the native structure has the minimum Gibbs 
   free energy.    
   If all chemical problems can be answered by quantum mechanics, there should be a 
   quantum mechanics derivation of Gibbs free energy formula 
   $G(\bX)$ for every possible conformation $\bX$
   of the protein.  We apply quantum statistics to derive such a formula.  
   For simplicity, only monomeric self folding globular proteins are covered. 
   
     We point out some immediate applications of the formula.  We
     show that the formula explains the observed phenomena very well.
     It gives a unified explanation to both folding and denaturation; it explains why
     hydrophobic effect is the driving force of protein folding and clarifies
     the role played by hydrogen bonding; it explains
     the successes and deficients of various surface area models.   
     The formula also gives a clear kinetic force of the folding: $\bF_i(\bX) = - \grad_{\bx_i}G(\bX)$.
     This also gives a natural way to perform the {\em ab initio} prediction of protein structure, minimizing
     $G(\bX)$ by Newton's fastest desciending method.

\end{abstract}
\section{Introduction}

    The newly synthesized peptide chain of a protein automatically 
   folds to its native structure and only in this native structure
   the protein can perform 
   its biological function.  Wrong structure will 
   cause disasters \cite{bt1999}. Why and how the protein folds to its native structure
   and how to predict the native structure from only the knowledge of the
   peptide chain are topics of protein folding \cite{dill2008}. 
        
       The fundamental law for protein folding is the {\bf Thermodynamic 
   Principle}:  the amino acid sequence of a protein determines its 
   native structure and the native structure of the protein
   has the minimum Gibbs free energy among all possible conformations
    \cite{anfinsen1973}.   Let $\bX$ be a conformation of a protein, is there
    a natural Gibbs free energy function $G(\bX)$?  The answer must be positive,
    as G.~N.~Lewis said in 1933: ``There are can be no doubt but that in quantum mechanics one 
    has the complete solution to the problems of chemistry."  (quoted from \cite[page 130]{bader1990}.)
    Protein folding is a problem in biochemistry, why we have not found such a formula $G(\bX)$?  The
    answer is also ready in hand.
    In 1929 Dirac wrote: ``The underlying physical laws necessary for the mathematical theory of ...
the whole of chemistry are thus completely known, and the difficulty is only that the exact application
of these laws leads to equations much too complicated to be soluble."  
(quoted from \cite[page 132]{bader1990}).   Yes, the complex of the Shr\"{o}dinger equation for
protein folding is beyond our ability to solve, no matter how fast and how powerful of our computers.   But
mathematical theory guarantees that there are a complete set of eigenvalues (energy levels) and eigenfunctions to the Shr\"{o}dinger equation in the Born-Oppenheimer approximation.   
Then consider that in the statistical mechanics, ensembles 
classify all (energy) states of the system, although we cannot have exact solutions to the Shr\"{o}dinger equation, we can apply the grand canonical ensemble to obtain the desired Gibbs free
energy formula $G(\bX)$.  This is the main idea of our derivation.   The interested readers
can read the details in the Appendix A.  

  Here we first state the formulae and the assumptions in deriving them.    Then we will
  point out some immediate applications and will use $G(\bX)$
  to explain well known facts
  such as hydrophobic effect and its relations with the hydrogen bonding, the denaturation of proteins,
  and the success in discriminating native and closely nearby compact non-native structures
  by empirical surface area models.   Other inferences from $G(\bX)$, such as the kinetic force
  in protein folding, the common practice of measuring $\lap G$, etc., are also discussed.  The 
  derivation itself will be put in Appendix A so that uninterested readers can skip it.   In Appendix B
  we will give the kinetic formulae $\bF_i(\bX)$.

  \subsection{Assumptions}
  
 All assumptions here are based on well-known facts of consensus.  Let $\fU$ be a protein with
 $M$ atoms $(\ba_1, \cdots, \ba_i, \cdots, \ba_M)$.   A structure of $\fU$ is
   a point $\bX = (\bx_1, \cdots, \bx_i, \cdots, \bx_M) \in \bbR^{3M}$,
   $\bx_i\in \bbR^3$ is the atomic center (nuclear) position of $\ba_i$.
   Alternatively, the conformation $\bX$ corresponds to a subset in $\bbR^3$,
   $P_\bX = \cup_{i=1}^MB(\bx_i, r_i) \subset \bbR^3$, where $r_i$ is the 
   van der Waals radius of the atom $\ba_i$ and $B(\bx, r) = 
   \{\by \in \bbR^3; |\by - \bx| \le r\}$ is a closed ball with radius $r$ and 
   center $\bx$.
 \begin{enumerate}
 \item
   The proteins discussed here are monomeric, single domain, self folding globular proteins.
  \item
   Therefore, in the case of our selected proteins, the environment of the protein folding, 
   the physiological environment, is pure water, there are no other
    elements in the environment, no chaperons, no co-factors, etc.   This is a rational simplification, 
    at least when one considers the environment as only the first hydration shell of a conformation,
    as in our derivation of the $G(\bX)$.
    \item 
    During the folding, the environment does not change.
    \item
     Anfinsen \cite{anfinsen1973} showed that before folding, the polypeptide chain already has
  its main chain and each residue's covalent bonds correctly formed.  Hence, our conformations should
  satisfy the following steric conditions set in \cite{fang2005} and \cite{fj2010}:
   there are $\epsilon _{ij} > 0$, $1 \le i < j \le M$ such
  that for any two atoms $\ba_i$ and $\ba_j$ in
   $P_\bX = \cup_{k=1}^M B(\bx_k, r_k)$,
  \begin{equation}\begin{array}{ll}
   \epsilon _{ij} \leq |\bx_i - \bx_j|, \!\!& \text{ no covalent bond between }\ba_i
   \text{ and } \ba_j;   \\
     d_{ij} - \epsilon_{ij} \leq |\bx_i - \bx_j|  \leq d_{ij} + \epsilon_{ij}, \quad
     & d_{ij}\text{ is the standard bond length between } \ba_i
      \text{ and } \ba_j. 
    \end{array}
   \label{steric1}
  \end{equation}
  We will denote all conformations satisfying (\ref{steric1}) as $\fX$ and only consider $\bX \in \fX$ in this paper.
  \item
  A water molecule is taking as a single particle, centered at
  $\bw \in \bbR^3$, the oxygen nuclear position, and the covalent bonds in it are
  fixed.   In the
  Born-Oppenheimer approximation, only the conformation $\bX$ is fixed, all
  particles, water molecules or electrons in the first hydration shell of $P_\bX$, are moving.
  \item
We agree that
simply classifying amino acids as hydrophobic or hydrophilic is an oversimplification \cite{eisenberg1986}.
All atoms should be classified according to the hydrophobicity of moieties or atom groups it belongs.  
Suppose there are $H$ hydrophobic levels $H_i$, $i = 1, \cdots, H$, such that
$\cup_{i=1}^HH_i = (\ba_1, \cdots, \ba_i, \cdots, \ba_M)$.
For example, we may assume that the classification is as in \cite{eisenberg1986}, there
are $H = 5$ classes, C, O/N, O$^-$, N$^+$, S.  Unlike in \cite{eisenberg1986},
we also classify every hydrogen atom into
one of the $H$ classes according to to whom it is bounded with.
There are many different
  hydrophobicity classifications.  Our derivation is valid
  for any of them.
\end{enumerate}

  \subsection{The Formula}
    
     The formula has
   two versions, the chemical balance version is:
 \begin{equation}
G(\bX) = \mu_eN_e(\bX) + \sum_{i=1}^H \mu_i N_i(\bX),
\label{jiben}
\end{equation}
where $N_e(\bX)$ is the mean number of electrons in the space included by 
the first hydration shell of $\bX$, $\mu_e$ is its chemical potential.
$N_i(\bX)$ is the mean number of water
molecules in the first hydration layer that directly contact to the atoms in $H_i$, 
$\mu_i$ is the chemical potential. 

  Let $M_\bX$ (see {\sc Figure 1}) be the molecular surface for the conformation $\bX$,
  defining $M_{\bX\, i}\subset M_\bX$ as the set of points in $M_\bX$ that are closer to atoms in 
  $H_i$ than any atoms in $H_j$, $j \ne i$.   Then the geometric version
  of $G(\bX)$ is:
\begin{equation}
  G(\bX) = a\mu_e V(\Omega_\bX) + ad_w\mu_eA(M_\bX) + 
  \sum_{i=1}^H\nu_i \mu_i A(M_{\bX_i}), \quad a, \nu_i > 0,
  \label{geo}
  \end{equation}
where $V(\Omega_\bX)$ is the volume of the domain $\Omega_\bX$ enclosed by $M_\bX$, $d_w$ the
diameter of a water molecule, and $A(M_\bX)$ and $A(M_{\bX_i})$ the areas of $M_\bX$ and $M_{\bX_i}$, $a[V(\Omega_\bX) + d_wA(M_\bX)] = N_e$, $\nu_i A(M_{\bX_i}) = N_i(\bX)$, $1\le i \le H$.   
The $a$ and $\nu_i$ are
independent of $\bX$, they are the average numbers of particles per unit volume and area.

\section{Applications}

\subsection{Structure Prediction}
Prediction of protein structures is the most important method to reveal proteins' functions and
working mechanics, it becomes a bottle neck in the rapidly developing life science.   
With more and more powerful computers, this problem is attacked in full
front.  Various models are used to achieve the goal, homologous or {\em ab initio}, full atom
model or coarse grained, with numerous parameters of which many are quite arbitrary.   
But although our computer power growths exponentialy, prediction power
does not follow that way.  At this moment, we should take a deep breath and remind what the
great physicist Fermi said: ``There are
two ways of doing calculations in theoretical
physics. One way, and this is the
way I prefer, is to have a clear physical picture
of the process that you are calculating. The
other way is to have a precise and selfconsistent
mathematical formalism." And ``I remember
my friend Johnny von Neumann used to
say, with four parameters I can fit an
elephant, and with five I can make him
wiggle his trunk."  Quoted from \cite{fermi}.

  These should also apply to any scientific calculation, not just theoretical physics.   Look at the current
  situation, all {\em ab initio} prediction models are actually just empirical with many parameters to
  ensure some success.      Fermi's comments remind us that a theory should be based on fundamental 
  physical laws, and contain no arbitrary parameters.  Look at formulae (\ref{jiben}) and (\ref{geo}), 
  we see immediately that
 they are neat, precise and self consistent mathematical formulae.  Furthermore, they
 including no arbitrary parameter, all terms in them have clear physical meanings.
 Chemical potentials $\mu_e$ and $\mu_i$'s, geometric constants $a$ and $\nu_i$'s, can be
 valued by theory or experiments, they are not arbitrary at all.

But a theory has to be developed, tested, until justified or falsified.  For interested researchers, the tasks
are to determine the correct values of the chemical potentials in (\ref{jiben})
and the geometric ratios $a$ and $\nu_i$ in (\ref{geo}).   There are many estimates to them,
but they are either for the solvent accessible surface area such as in \cite{eisenberg1986} hence
not suit to the experiment data as pointed out in \cite{tsp1992}, or do not distinguish different
hydrophobicity levels as in \cite{tsp1992}.   To get the correct values of the chemical potentials and geometric constants, commonly
  used method of training with data can be employed, in which we can also test the formulae's ability
of discriminating native and nearby compact non-native structures.
After that, a direct test is to
predict the native structure from the amino acid sequence of a protein by minimizing the following:
\begin{equation}
   G(\bX_N) = \inf_{\bX \in \fX}G(\bX).
 \label{mini}
 \end{equation}
This is the first time that we have a theoretically derived formula of the Gibbs free energy.   Before
this, all {\em ab intitio} predictions are not really {\em ab initio}.     A combined (theoretical and experimental)
search for the values of chemical potentials will be the key for the success of the {\em ab initio}
prediction of protein structure.

\subsection{Energy Surface or Landscape}
An obvious application is the construction of Gibbs free energy surface or landscape.   We do not
need any empirical estimate anymore, the Gibbs free energy formula $G: \fX \to \bbR$ gives
a graph $(\bX, G(\bX))$ over the space $\fX$ (all eligible conformations for a 
given protein), and this is nothing but the Gibbs free energy surface.  
Mathematically it is a $3M$ dimensional hyper-surface.   Its characteristics concerned by
students of energy surface theory, such as how rugged it is, how many local minimals are there, is there a
funnel, etc., can be answered by simple calculations of the formula.

  Since the function $G$ is actually defined on the whole $\bbR^{3M}$ (on an domain of $\bbR^{3M}$ 
  containing all $\fX$ is enough),  we can explore mathematical tools to study its graph, and compare
  the results with the restricted conformations.   One important question is:  Does the absolute minimum
  structure belongs to $\fX$?

\subsection{Kinetics}
It is observed that while we apply the thermodynamic principle, a difficulty is that we
do not have kinetics and have use other method to present it \cite{hyeon2010}.  
The advantage of a theoretical formula for Gibbs free energy in the form $G(\bX)$
is that it connects the thermodynamics
with the kinetics.  In fact, for any atomic position $\bx_i$, the kinetic force is 
$\bF_i(\bX) = - \grad_{\bx_i}G(\bX)$, \cite{dai2007}.  With formula (\ref{geo}) these
quantities are really calculable, mathematical formulae and implementations on
molecular surface $M_\bX$ are given in \cite{fj2008}.
We will give the mathematical formulae in Appendix B.   The resulting Newton's fastest descending 
method was used in the simulation in \cite{fj2010}.

\section{Discussions}
  
   We are theoretically treating the protein folding by introducing quantum statistics.
  A theory is useful only if it can make explanations to the observed facts and if it can
  simplify and improve research methods as well as clarify concepts.   We will show
  that $G(\bX)$ can do exactly these.


       \begin{center}
 \scalebox{3.8}[3.8] 
 {\resizebox{!}{100pt}     
{\includegraphics{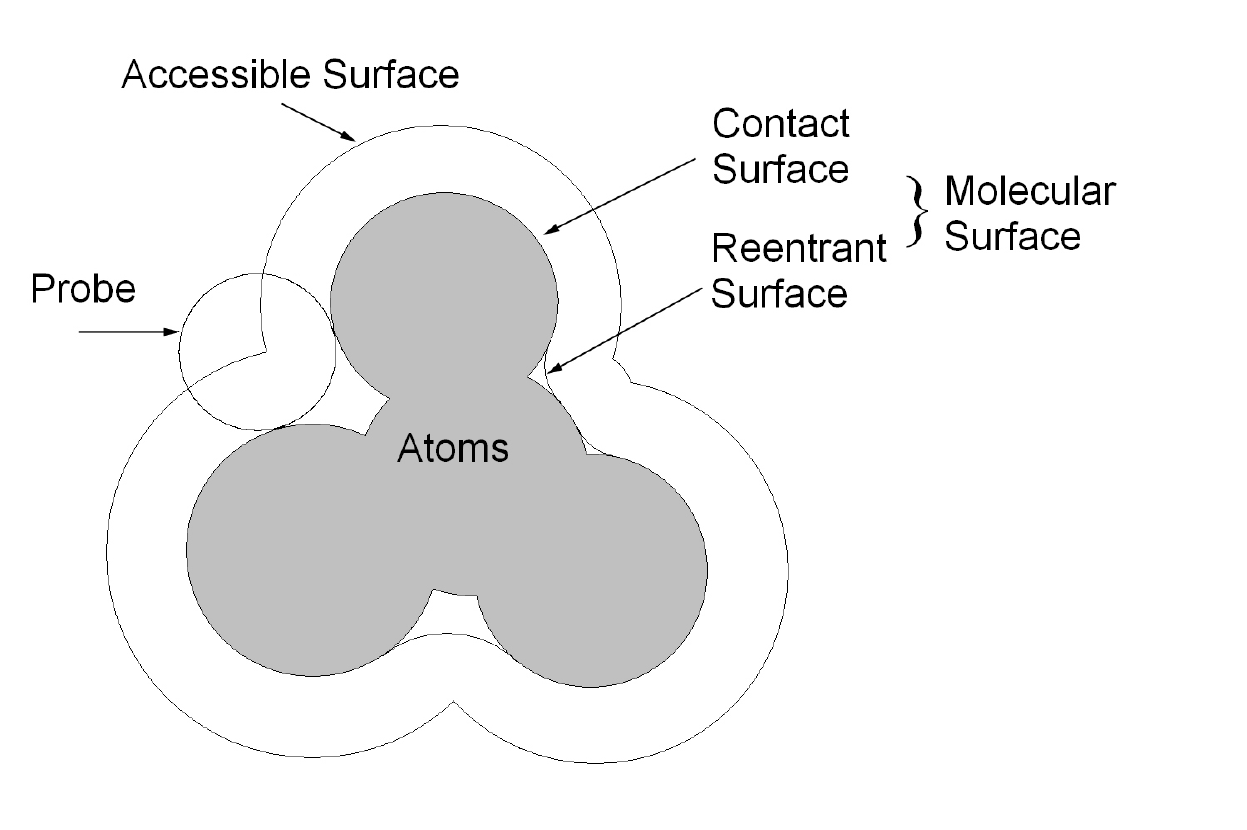}}}
\end{center}

\vspace{-1.0in}

 {\sc Figure 1.} Two dimensional presenting of molecular surface \cite{richards1977}
 and solvent accessible surface \cite{lr1971}.  This figure was originally in \cite{fj2010}.



       \begin{center}
 \scalebox{3.9}[3.9] 
 {\resizebox{!}{100pt}     
{\includegraphics{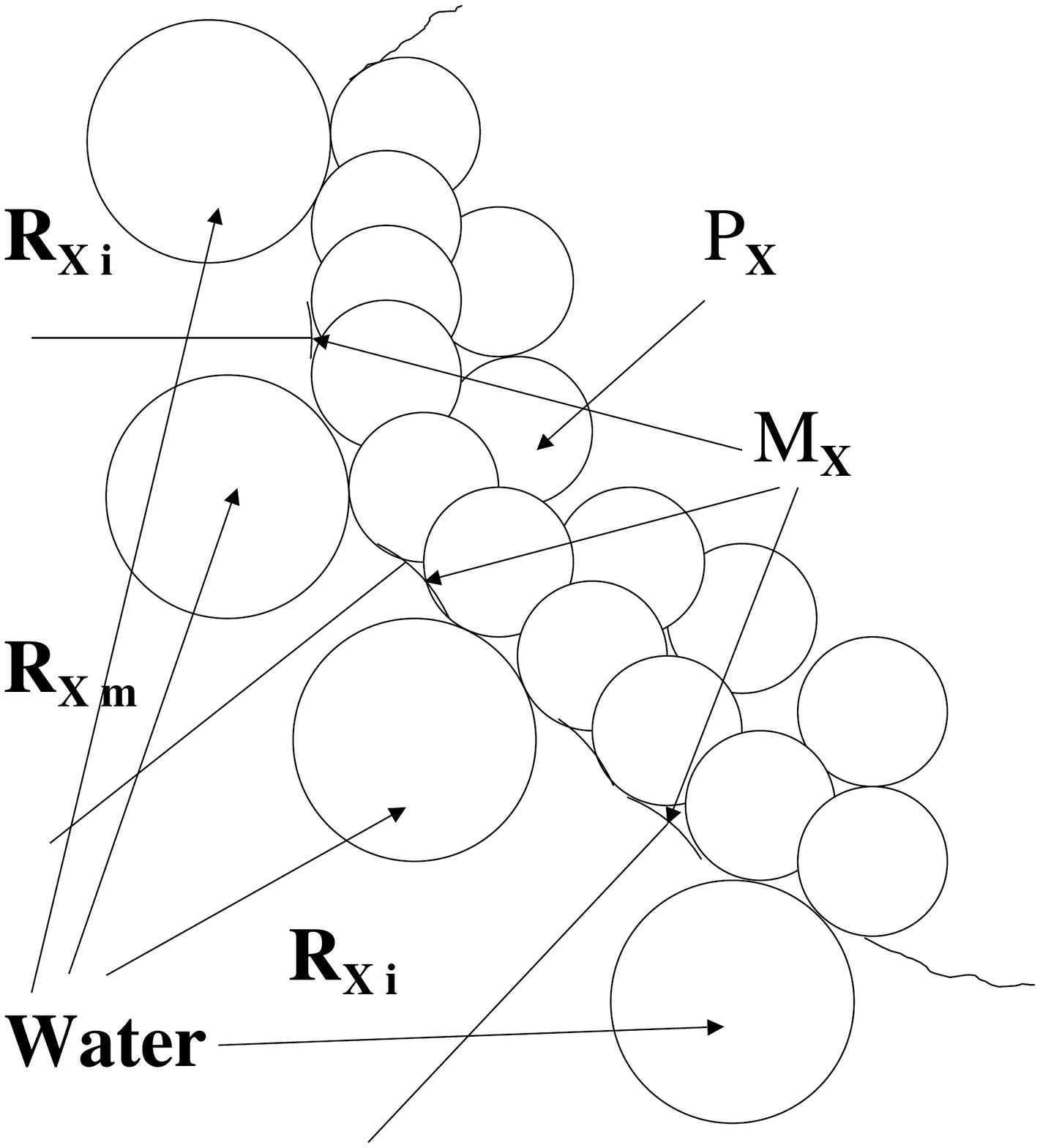}}}
\vspace{-0.8in}

 {\sc Figure 2.} Note that $\cR_{\bX i}$ generally are not connected, i.e., having more than one block.
\end{center}


   If the same theoretical result can be derived from two different disciplines, it is often not just by
   chance.  We will show an early phenomenological mathematical model \cite{fang2005},
   starting from purely geometric reasoning, has achieved formula (\ref{geo}), with just
   two hydrophobic levels, hydrophobic and hydrophilic.
  
  A theory also
  has to be falciable, that is making a prediction to be checked.   The fundamental prediction
  is that minimizing formulae (\ref{jiben}) or (\ref{geo}) we will get the native structures from the
  amino acid sequences of proteins covered in the assumptions of the formulae.    
   That can only be done after we have the actual values in the physiological
  environment of the chemical potentials appear in the formulae.

\subsection{Unified Explanation of Folding and Denaturation}

  Protein denaturation is easy to happen, enve if the environment is slightly changed,
  as described in \cite{wuxian1931}
  by Hsien Wu in 1931.  (The reference \cite{wuxian1931} is the 13th article that theorizes the results  of a 
  series experiments, and a preliminary report was read before the Xlllth International Congress of Physiology
  at Boston, August 19-24, 1929, and published in the {\em Am. J. Physiol} for October 1929.
  In which Hsien Wu first suggested that the denatured protein is still the same molecule, only structure
  has been changed.)  Anfinsen in various experiments showed that after denaturation
  by changed environment, if removing the denature agent, certain globular proteins can 
  spontaneously refold to its native
  structure, \cite{anfinsen1973}.   The spontaneous renaturation suggests that protein folding does not need
  outside help, at least to the class of proteins in study.   Therefore, the fundamental law of thermodynamics
  asserts that in the environment such that a protein can fold, the native structure must have the
  minimum Gibbs free energy.   The same is true for denaturation, under the denatured environment,
  the native structure no longer has the minimum Gibbs free energy, some other structure(s), will 
  have the minimum Gibbs free energy.  Thus let $En$ present environment, any formula of Gibbs free 
  energy should be stated as $G(\bX, En)$ instead of just $G(\bX)$, unless specified the environment like
  in this paper.   Let $En_N$ be the physiological
  environment and $En_U$ be some denatured environment, $\bX_N$ be the native structure and
  $\bX_U$ be one of the denatured stable structure in $En_U$, then the
  thermodynamic principle for both of protein folding and unfolding should be that
  \begin{equation}
    G(\bX_N, En_N) < G(\bX_U, En_N), \quad G(\bX_N, En_U) > G(\bX_U, En_U).  
 \label{twoway}
  \end{equation}
  To check this, an experiment should be designed that can suddenly put proteins in a different environment.
  Formulae (\ref{jiben}) and (\ref{geo}) should be written as $G(\bX, En_N)$.  Indeed,
 the chemical potentials $\mu_e$ and $\mu_i$'s are Gibbs free energies per corresponding particles,
 $\mu = u + Pv - Ts$.   Two environment parameters, temperature $T$ and pressure $P$, explicitly
 appear in $\mu$, the inner energy $u$ and entropy $s$ may also implicitly depend on the environment.   According
 to formulae (\ref{jiben}) and (\ref{geo}), if $\mu_i < 0$, then make more $H_i$ atoms to expose to water 
 (make larger $A(M_{\bX\, i})$) will reduce the Gibbs free energy.  If $\mu_i > 0$, then the reverse will 
 happen.  Increase or reduce the $H_i$ atoms' exposure to water ($A(M_{\bX\, i})$), the
 conformation has to change.  The conformation changes to
 adjust until we get a conformation $\bX_N$, such that the net effect of
 any change of it will either increase
 some $H_i$ atoms' exposure to water while $\mu_i >0$ or reduce $H_i$ atoms' exposure to water
 while $\mu_i < 0$.  In other words, the $G(\bX, En_N)$ achieves its minimum at $G(\bX_N, En_N)$.   
 Protein folding, at least for the proteins considered in the assumptions, is explained very well by 
 formulae $(\ref{jiben})$ and (\ref{geo}).

 In changed environment, the chemical potentials $\mu_e$ and $\mu_i$'s in formulae 
 (\ref{jiben}) and (\ref{geo}) changed their values.  With the changed
 chemical potentials, $G(\bX, En_U)$ has the same form as $G(\bX, En_N)$ but different chemical
 potentials.  Therefore, the structure $\bX_U$ will be stable, according to the second inequality in (\ref{twoway}), the process is exactly the same as described for the protein folding if
  the changing environment method does not include introducing new kinds (non-water) of particles, 
  for example, if we only change temperature or pressure.
 
 Even the new environment including new kinds of particles, formulae (\ref{jiben}) and (\ref{geo})
 can still partially explain the denaturation, only that more obstructs prevent the protein to denature
 to $\bX_U$, but any way it will end in some structure other than the $\bX_N$, the protein is denatured.   
 Actually, this is a hint
 of how to modify the current formulae to extend to general proteins.
 
 \subsection{Why $G(\bX)$ Instead Of $\lap G(\bX)$}
   Here is a chance to explain why we use $G(\bX)$ instead of $\lap G(\bX)$.
   In various experiments of testing the Gibbs free energy difference ($\lap G$) between 
   $\bX_N$ and $\bX_U$, 
   the common practice is essentially set 
   \begin{equation}
     \lap G = G(\bX_U, En_U) - G(\bX_N, En_N). 
    \label{cuo}
   \end{equation}
   see, for example, \cite{cooper}.   Though some interpolation was taken to adjust, but that
   is not the experiment observation.
   But formulae (\ref{jiben}) and (\ref{geo}) suggest that what we need is
      \begin{equation}
     \lap G = G(\bX_U, En_N) - G(\bX_N, En_N).
     \label{environ}
     \end{equation}
     Unfortunately, there is no method of denaturation 
     without changing environment, at least currently no such method. 
     Therefore, no way to experimentally
     measuring of $\lap G$ in (\ref{environ}).   We should reexamine
     the conclusion of $\lap G$ is very small because it was essentially drawn from (\ref{cuo}).
     Thus although we believe that it is true, the conclusion was achieved neither via theory nor 
     via real experiment observation.
     
       While experiment has no way to change the native structure without disturbing the environment,
       theory can play a role instead.   Formulae (\ref{jiben}) and (\ref{geo}) give us the chance
       to compare $\lap G$, as long as we have accurate chemical potentials.

  \subsection{Explain Hydrophobic Effect and the Role Played by Hydrogen Bonding}
 In 1959, by reviewing the literature Kauzmann concluded that the hydrophobic effect is the main 
driving force in protein folding \cite{kauzmann1959} .    Empirical correlation between hydrophobic free energy and aqueous cavity surface area was noted as early as 1974 \cite{tanford1974}, giving
justification of the hydrophobic effect.  Various justifications of hydrophobic effect were published,
based on empirical models of protein folding, for example, \cite{dill1990}.   
But the debate continues to present, some still insist that it is the hydrogen bond instead of hydrophobic
effect plays the main role of driving force in protein folding, for example, \cite{rose2006}.   The theoretically
derived formulae (\ref{jiben}) and (\ref{geo}) can explain why the hydrophobic effect is indeed the
driving force.  A simulation of reducing hydrophobic area alone (\cite{fj2010}) can
explain the intra-molecular hydrogen bonds.

  In fact,
  according to formulae (\ref{jiben}) and (\ref{geo}), if $\mu_i < 0$, then make more $H_i$
  atoms to appear in the boundary of $P_\bX$ will reduce the Gibbs free
 energy.  If $\mu_i > 0$, then the reverse will happen, reducing the exposure of
 $H_i$ atoms to water will reduce the Gibbs free energy.   This gives a theoretical explanation
 of the hydrophobic effect.  The kinetic formulae $\bF_i = \grad_{\bx_i}G(\bX)$ and
 those given in Appendix B are the force that push the conformation to change to the native structure.
 
 The mechanics stated above works through the chemical potentials $\mu_i$ for various levels of 
 hydrophobicity, in physiological environment,
 all hydrophobic $H_i$'s will have positive $\mu_i$, all hydrophilic $H_i$'s will have negative
 $\mu_i$.   Thus changing conformation $P_\bX$ such that the most hydrophilic $H_i$
 ($\mu_i = \min (\mu_1, \cdots, \mu_H$) gets the first priority to appear on the boundary, 
 and the most hydrophobic $H_i$ ($\mu_i = \max (\mu_1, \cdots, \mu_H)$) gets the first priority to
 hide in the hydrophobic core to avoid contacting with water
 molecules, etc.   We should keep in mind that all the time, the steric conditions (\ref{steric1}) have
 to be obeyed.

  But the hydrophobic effect is actually partially working through hydrogen bond formation.   
  This is well presented in
  the chemical potentials in (\ref{jiben}) and (\ref{geo}).   In fact, the values of the chemical
  potentials reflect the ability of the atoms or atom groups to form hydrogen bond, either with
  another atom group in the protein or with water molecules.   This gives a way to theoretically
  or experimentally determine the values of hydrophilic chemical potentials: checking the actually
  energy of the hydrogen bond.  
  
  For hydrophobic ones, it will be more complicated, common sense
  is that it reduces the entropy that certainly comes from the inability of forming hydrogen bonds with
  water molecules.   Hence although hydrophobic effect is the driving force of protein folding, it
  works through the atom's ability or inability to form hydrogen bonds with water molecules.

  How to explain the intra-molecular hydrogen bonds?   It seems that formula (\ref{jiben}) and (\ref{geo})
  do not address this issue.   The possible theory is that the amino acid sequence of a protein is highly
  selectable in evolution, in tact only a tiny number of amino acid sequences can really become a
  protein.   With these specially selected sequences, while shrinking the various hydrophobic surfaces to
  form a hydrophobic core, residues are put in position to form secondary structures and their associated
  hydrogen bonds.   This sounds a little bit too arbitrary.  But a simulation of shrinking hydrophobic surface
  area alone indeed produced secondary structures and hydrogen bonds.   The simulation was reported
  in \cite{fj2010}.   Without calculating any dihedral angles or electronic charges, without any 
  arbitrary parameter, paying no attention to any particular atom's position, 
  by just reducing hydrophobic surface 
  area (there it was assumed that there are 
  only two kinds of atoms, hydrophobic and hydrophilic), secondary structures and hydrogen bonds duly
  appeared.  The proteins used in the simulation are 2i9c, 2hng, and 2ib0, with 123, 127, and 162 residues. 
  No simulation of any kind of empirical or theoretical models had achieved such a success.    More than
  anything, this simulation should prove that hydrophobic effect alone will give more chance of forming
  intra-molecular hydrogen bonds.   Indeed, pushing hydrophilic atoms to make hydrogen bonds
  with water molecules will give other non-boundary hydrophilic groups more chance to form
  intra-molecular hydrogen bonds.

   Again formula (\ref{geo}) can partly explain the success of this simulation, when there are only two 
   hydrophobic classes in (\ref{geo}), the hydrophobic area presents the main positive part of the 
   Gibbs free energy, reducing it is reducing the Gibbs free energy, no matter what is the chemical 
   potential's real value.

  \subsection{Explanation of the Successes of Surface Area Models}

In 1995, Wang {\em et al} \cite{wang1995}
compared 8 empirical energy models by testing their ability
to distinguish native structures and their close neighboring compact non-native structures.   Their models 
WZS are accessible surface area models with 14 classes of atoms, 
$\sum_{i=1}^{14}\sigma_i A_i$.  Each two combination of three targeting proteins were used to train 
WZS to get $\sigma_i$, 
hence there are three models WZS1, WZS2, and WZS3.   Among the 8 models, all WZS's performed the best, distinguishing
all 6 targeting proteins.  The worst performer is the force field AMBER 4.0, it failed in distinguishing any of
the 6 targets.  
            
These testing and the successes of various surface area models such as \cite{eisenberg1986}, showed
that instead of watching numerous pairwise atomic interactions, the surface area models, though
looking too simple, have surprising powers.   Now the formula (\ref{geo}) gives them a theoretic  justification.
On the other hand, the successes of these models also reenforce the theoretical
results.

   There is a gap between the accessible surface area model in \cite{eisenberg1986} and the experiment
   results (surface tension), as pointed out in \cite{tsp1992}.   
   The gap disappeared when one uses the molecular surface area
   to replace the accessible surface area, in \cite{tsp1992} it was shown that molecular surface area
assigned of 72-73 cal/mol/\AA$^2$ perfectly fits with the macroscopic experiment data.      Later it was asserted that the molecular surface
   is the real boundary of protein in its native structure \cite{jackson1993}.
   
     {\sc Figure 1} and {\sc Figure 2} show the water molecules contact to $P_\bX$ and
     the accessible surface and molecular surface, we see that all water molecules must be outside
     the molecular surface $M_\bX$, but the assessable surface is in the middle of the first hydration shell.
     So it is better to use the molecular surface $M_\bX$ as the boundary of the conformation $P_\bX$. 
     Moreover, the conversion of the mean
     numbers $N_i(\bX)$ to surface area, $N_i(\bX) = \nu_iA(M_{\bX\,. i})$, only works for the molecular surface,
     not for the accessible surface.   This can explain the conclusions  in \cite{tsp1992} and \cite{jackson1993}.

In fact, the advantage of the solvent accessible surface is that by definition of it we know exactly each atom
occupies which part of the surface, therefore, one can calculate its share in surface area.  
This fact may partly account why there are so many models based on the solvent accessible
surface, even people knew the afore mentioned gap.  For other surfaces, we have to define the part of surface that  
belongs to a specific hydrophobicity class.   This was resolved in \cite{fang2005} via the distance function
definition as we used here.

All surface area models neglected one element, the volume of the structure.   As early as
in the 1970's, Richards and his colleagues already pointed out that the native structure of globular 
proteins
is very dense, or compact, (density $= 0.75$, \cite{richards1977}).   To make a conformation denser, 
obviously we
should shrink the volume $V(\Omega_\bX)$.  The model in \cite{fang2005} introduced
volume term but kept the oversimplification of all atoms are either hydrophobic or
hydrophilic.   The derivation of (\ref{jiben}) and (\ref{geo}) shows that volume term should be counted,
but it may be that $a\mu_e$ is very small, in that case, volume maybe really is irrelevant.

\subsection{Coincidence with Phenomenological Mathematical Model}
If a theoretical result can be derived from two different disciplines, its possibility of correctness
will be dramatically increased.   Indeed, from a pure geometric consideration, a phenomenological
mathematical model, $G(\bX) = aV(\Omega_\bX) + bA(M_\bX) + cA(M_{\bX \,1})$, $a, b, c > 0$ 
(it was assumed that there are only two hydrophobicity levels, hydrophobic and hydrophilic, the
hydrophilic surface area $A(M_{\bX\, 2})$ is absorbed in $A(M_\bX)$ by $A(M_{\bX\, 2})
= A(M_\bX) - A(M_{\bX\, 1}))$,
was created in \cite{fang2005}.   It was based on the well-known global geometric
characteristics of the native structure of globular proteins: 1. high density; 2. smaller surface area;
3. hydrophobic core, as demonstrated and summarized in \cite{richards1977} and
\cite{novotny1984}.  So that to obtain the native structure, we should shrink the volume (increasing
the density) and surface area, and form better hydrophobic core (reducing the hydrophobic
surface area $A(M_{\bX\, 1})$) simultaneously and cohensively.

  The coincidence of formula (\ref{geo}) and the phenomenological mathematical model of 
  \cite{fang2005} cannot be just a coincidence.  Most likely, it is the same natural law reflected in 
  different disciplines.   The advantage of (\ref{geo}) is that everything there has its physical meaning.


 \subsection{Potential Energy Plays No Role in Protein Folding}
 Formulae (\ref{jiben}) and (\ref{geo}) theoretically show that hydrophobic effect is the driving force of 
  protein folding, it is
  not just solvent free energy besides the pairwise interactions such as the Coulombs, etc., as all force fields 
  assumed.  
  Only in the physiological environment the hydrophobic effect works towards to native structure, otherwise 
  it will push
  denaturation as discussed in explanation of folding and unfolding.  
 Formulae (\ref{jiben}) and (\ref{geo}) show that the Gibbs free energy
 is actually independent of the potential energy, against one's intuition and a bit of surprising.   The explanation
 is that during the folding process, all covalent bonds in the main chain and each side chain are kept invariant,
the potential energy has already played its role in the synthesis process of forming the peptide chain,
 which of course can also be described by quantum mechanics.   According to Anfinsen \cite{anfinsen1973},
 protein folding is after the synthesis of the whole peptide chain, so we can skip the synthesis process
 and concentrate on the folding process.  
 
  The steric conditions (\ref{steric1}) will just keep this early synthesis result, not any $\bX = (\bx_1,
  \cdots, \bx_i, \cdots, \bx_M)$ is eligible to be a conformation, it has to satisfy (\ref{steric1}).
   The steric conditions not only
  pay respect to the bond length, it also reflect a lot of physi-chemco properties of a conformation:
  They are defined
   via the allowed minimal atomic distances,  such that for non-bonding
     atoms, the allowed minimal distances are: 
     shorter between differently charged or polarized atoms; 
     a little longer between non-polar ones; and much longer (generally greater than the sum of their radii)
     between the same charged  ones, etc.  For example, we allow minimal distance
     between sulfur atoms in Cysteines to form disulfide bonds.  And for any new found
    intra-covalent bond between side chains, we can easily modify the steric conditions to allow
    it to form during folding, though it may not necessarily form.
    
     Especially in the minimization of $G(\bX)$, steric conditions must be kept, thus the minimization
     in (\ref{mini}) is
     a constrained minimization.   This, unfortunately, is a draw back, it increased the mathematical
     difficulty.

\section{Conclusion}
  A quantum statistical theory of protein folding for monomeric, single domain, self folding globular proteins
  is suggested.  The assumptions of the theory fit all observed realities of protein folding.   The resulting
  formulae (\ref{jiben}) and (\ref{geo}) do not have any arbitrary parameters and all terms
  in them have clear physical meaning.  Potential energies involving pairwise interactions between
  atoms do not appear in them.   
  
  Formulae (\ref{jiben}) and (\ref{geo}) have explanation powers.  They give unified explanation to folding 
  and denaturation, to
  the hydrophobic effect in protein folding and its relation with the hydrogen bonding.    The formulae
  also explain the relative successes of surface area protein folding models.  Relation between
  kinetic and thermodynamic of protein folding is discussed, driving force formula comes from the 
  Gibbs free energy formula (\ref{geo}) are also given.   Energy surface theory
  will be much easier to handle.  The concept of $\lap G$ is clarified. 
 
 \appendix
 
\section{The Derivation}

   Let $d_w$ be the diameter of a water molecule and
     $M_\bX$ be the molecular surface of $P_\bX$ as defined in \cite{richards1977} with
    the probe radius $d_w/2$, see {\sc Figure 1}.
Define 
  \begin{equation}
  \cR_\bX = \{\bx \in \bbR^3:\; \dist(\bx, M_\bX) \le d_w\} \setminus P_\bX
  \label{huan}
  \end{equation}

  \n as the first hydration shell surrounding $P_\bX$, where $\dist(\bx, S) = \inf_{\by \in S}|\bx - \by|$.  
  Then $\cT_\bX = P_\bX \cup \cR_\bX$ will
  be our thermodynamic system of protein folding at the conformation $\bX$.

  We classify the atoms in $\fU$ into $H$ hydrophobicity classes
  $H_i$, $i = 1, \cdots, H$, such that $\cup_{i=1}^H H_i = \{\ba_1, \ba_2, \cdots, \ba_M\}$.
   Let $I_i \subset \{1, 2, \cdots, M\}$ be the subset such that $\ba_j \in H_i$ if and only if $j\in I_i$.  Define
  $P_{\bX \, i} = \cup_{j\in I_i}B(\bx_j, r_j) \subset P_\bX$ and as shown in {\sc Figure 2},
  \begin{equation}
  \cR_{\bX\, i} = \{\bx \in \cR_\bX:\; \dist(\bx, P_{\bX_i}) \le \dist(\bx, P_\bX \! \setminus \!
  P_{\bX \, i})\}, \quad 1 \le i \le H,
  \label{fenhuan}
  \end{equation}

  Let $V(\Omega)$ be the volume of $\Omega \subset \bbR^3$, then
 \begin{equation}
 \cR_\bX = \cup_{i=1}^H\cR_{\bX \, i}, \quad V(\cR_\bX) = \sum_{i=1}^HV(\cR_{\bX\, i}),
\text{  and for } i\ne j, \quad V(\cR_{\bX \, i} \cap \cR_{\bX \, j}) = 0.
 \label{fenhao}
 \end{equation}

  Since $M_\bX$ is a closed surface, it divides $\bbR^3$ into two regions
  $\Omega_\bX$ and $\Omega'_\bX$ such that 
  $\bdy \Omega_\bX = \bdy \Omega'_\bX =
  M_\bX$ and $\bbR^3 = \Omega_\bX \cup M_\bX \cup \Omega'_\bX$.   We have
  $P_\bX \subset \Omega_\bX$
  and all nuclear centers of atoms in the water molecules in $\cR_\bX$ are contained in $\Omega'_\bX$. 
  Moreover, $\Omega_\bX$ is
  bounded, therefore, has a volume $V(\Omega_\bX)$.
   Define the hydrophobicity subsurface $M_{\bX\, i}$, $1\le i \le H$, as
  \begin{equation}
    M_{\bX\, i} = M_\bX \cap \overline{\cR_{\bX\, i}}.
  \label{fenmian}
  \end{equation}
  Let $A(S)$ be the area of a surface $S\subset \bbR^3$, then
  \begin{equation}
  M_\bX = \cup_{i=1}^H M_{\bX\, i}, \quad A(M_\bX) = \sum_{i=1}^H A(M_{\bX \, i}), \text{      and if }i \ne j, \text{ then } 
  A(M_{\bX \, i}\cap M_{\bX \,j}) = 0.
  \label{mianjiao}
  \end{equation}

         Although the shape of each atom in $\fU$ 
  is well defined by the theory of atoms in molecules (\cite{bader1990} and
  \cite{popelier2000}),
   what concerning us here is the overall shape of the structure $P_\bX$.
  The cutoff of electron density $\rho \ge 0.001$au (\cite{bader1990} 
  and \cite{popelier2000}), gives the overall
  shape of a molecular structure that is just like $P_\bX$,
  a bunch of overlapping balls.   Moreover, the boundary of the $\rho \ge 0.001$au
  cutoff is much similar to the {\bf molecular surface} $M_\bX$ which was defined by
  Richards in 1977 \cite{richards1977} and was shown has more physical meaning
  as the boundary surface of the conformation $P_\bX$ \cite{tsp1992} and \cite{jackson1993}.
 
\subsection{The Shr\"{o}dinger Equation} 

  For any conformation $\bX \in \fX$, let $\bW = (\bw_1, \cdots, \bw_i, \cdots, \bw_N) \in \bbR^{3N}$
  be the nuclear centers of water molecules in $\cR_\bX$ and
  $\bE = (\be_1, \cdots, \be_i, \cdots, \be_L)\in \bbR^{3L}$ be electronic
  positions of all electrons in $\cT_\bX$.    Then the
   Hamiltonian for the system $\cT_\bX$ is
    \begin{equation}
   \hat H  =  \hat T + \hat V = -\sum_{i=1}^M\frac{\hbar^2}{2m_i} \grad^2_i 
   - \frac{\hbar^2}{2m_w}\sum_{i = 1}^{N}
    \grad^2_i  -  \frac{\hbar^2}{2m_e} \sum_{i=1}^{L}\grad^2_i 
  +  \hat V(\bX, \bW, \bE),
 \end{equation}
 where $m_i$ is the nuclear mass of atom $\ba_i$ in $\fU$, $m_w$ and 
 $m_e$ are the masses of water molecule and electron;
  $\grad^2_i$ is Laplacian in corresponding $\bbR^3$; and $V$ the potential. 
\vspace{0.1in}

\subsection{The First Step of The Born-Oppenheimer Approximation}

Depending on the shape of $P_\bX$, for each $i$, $1\le i \le H$, the maximum numbers 
$N_{\bX \, i}$ of water molecules contained in 
$\cR_{\bX \, i}$ vary.  Theoretically we consider all cases, i.e., 
there are $0 \le N_i \le N_{\bX \, i}$ water 
molecules in $\cR_{\bX \, i}$, $1 \le i \le H$.
Let $M_0 = 0$ and $M_i = \sum_{j\le i}N_j$ and
 $\bW_i = (\bw_{M_{i -1} + 1}, \cdots, \bw_{M_{i-1} + j}, \cdots,
  \bw_{M_i}) \in \bbR ^{3N_i}$, $1\le i \le H$, and $\bW = (\bW_1, \bW_2, \cdots, \bW_{M_H})
 \in \bbR^{3M_H}$ denote the nuclear positions of water molecules in $\cR_\bX$.
     As well, there will be all possible numbers $0 \le N_e < \infty $ of electrons in $\cT_\bX$.   Let
 $\bE = (\be_1, \be_2, \cdots, \be_{N_e}) \in \bbR^{3N_e}$ denote their nuclear positions.
For each fixed $\bX \in \fX$ and  $N = (N_1, \cdots, N_H, N_e)$, the Born-Oppenheimer approximation has
 the Hamiltonian
   \begin{equation*}
 \hat H_X  =  - \frac{\hbar ^2}2\left \{ \frac1{m_w}\sum_{j=1}^{M_H}
  \grad^2_j +  \frac1{m_e}\sum_{\nu =1}^{N_e} \grad^2_\nu  \right \} 
     +   \hat V(\bX, \bW, \bE).
  \end{equation*}
The eigenfunctions $\psi_i^{\bX, N} (\bW,\bE) 
 \in L^2_0(\prod_{i=1}^H\cR_{\bX \, i}^{N_i} \times \cT_\bX^{N_e})  = \cH_{\bX, N}$,
  $1\le i < \infty $, comprise an orthonormal basis of $\cH_{\bX, N}$.  Denote
 theire eigenvalues (energy levels) as $E^i_{\bX, N}$, 
 then $\hat H_\bX \psi_i^{\bX, N} = E^i_{\bX, N} \psi_i^{\bX, N}$.
  
\subsection{Grand Partition Function and Grand Canonic Density Operator}
 
 In the following we will use the natotions and definitions in \cite[Chapter 10]{greiner1994}.
 Let $k_B$ be the Bolzmman constant, set  $\beta = 1/k_BT$.   Since the numbers 
 $N_i$ and $N_e$ vary, we should adopt the grand canonic ensemble.   
 Let $\mu_i$ be the chemical potentials, that is,
 the Gibbs free energy per water molecule in $\cR_{\bX \, i}$. Let
 $\mu_e$ be electron chemical potential.
  The grand canonic density operator is
 (\cite{greiner1994} and \cite{dai2007})
 \begin{equation*}
 \hat \rho_\bX
   = \exp \left \{-\beta \left [ \hat H_\bX - \sum_{i=1}^H\mu_i
   \hat N_i  - \mu_e \hat N_e - \Omega(\bX) \right ] \right \}.
 \end{equation*}
 where the grand partition function is
 \begin{eqnarray*}
 \exp[-\beta \Omega(\bX)] & = &
  \text{Trace}\left \{\exp \left [-\beta \left (\hat H_\bX - \sum_{i=1}^H\mu_i \hat N_i
    - \mu_e \hat N_e \right )\right ] \right \} \\
  & = & \sum_{i, N}e^{-\beta [E^i_{\bX, N} - \sum_{i=1}^H\mu_i N_i
   - \mu_e N_e]}.
  \label{miao}
 \end{eqnarray*}

 \subsection{The Gibbs Free Energy $G(\bX)$}  
 
 According to \cite[page 273]{greiner1994}, under the grand canonic ensemble
 the entropy $S(\bX) = S(\cT_\bX)$ of the system $\cT_\bX$ is
 \begin{eqnarray}
  S(\bX) & = & - k_B \text{Trace}(\hat \rho_\bX \ln \hat \rho_\bX) = -k_B \langle \, \ln \hat \rho_\bX \, \rangle 
   =   k_B \beta \left \langle \hat H_\bX - \Omega(\bX) - \sum_{i=1}^H\mu_i 
   \hat N_i -  \mu_e \hat N_e \right \rangle  \nonumber \\
   & = &  \left [ \langle \hat H_\bX \rangle - \langle \Omega(\bX) \rangle
   - \sum_{i=1}^H  \mu_i\langle \hat N_i \rangle 
   - \mu_e \langle \hat N_e \rangle  \right ] / T \nonumber \\
   &  =  & \left [ U(\bX) - \Omega(\bX) 
   - \sum_{i=1}^H  \mu_iN_i(\bX) - \mu_e N_e(\bX)  \right ] / T.
 \label{entropy}
 \end{eqnarray}  
 Here we denote $\langle \hat N_i \rangle
 = N_i(\bX)$ the mean numbers of water molecules  
 in $\cR_{\bX \, i}$, $1\le i \le H$, and $\langle \hat N_e \rangle =
 N_e(\bX)$ the mean number of electrons in $\cT_\bX$.  
 The inner energy  $\langle \hat H_\bX \rangle $ of the system
 $\cT_\bX$ is denoted as $U(\bX) = U(\cT_\bX)$. 
  The term $\Omega(\bX)$ is a state function with variables $T, V, \mu_1, \cdots, \mu_H$, and $ \mu_e$,
  and  is called the grand canonic potential (\cite[page 27]{greiner1994}) or the thermodynamic potential
  (\cite[page 33]{dai2007}).  By the general thermodynamic equations 
  \cite[pages 5 and 6]{dai2007}:
 \begin{equation*}
 d\Omega(\bX) = - SdT - PdV - \sum_{i=1}^HN_i d\mu_i  - 
 N_e d\mu_e,  \quad
  \lambda 
 \Omega(\bX) = \Omega(\bX) (T, \lambda V, \mu_1,  \cdots,  \mu_H, \mu_e),
 \end{equation*}
 we see that  $\Omega(\bX)(T, V, \mu_1, \cdots, \mu_H, \mu_e) = - PV(\bX)$, where
 $V(\bX) = V(\cT_\bX)$ is the volume of the thermodynamic system $\cT_\bX$.
Thus by (\ref{entropy}) we obtain the Gibbs free energy $G(\bX) = 
G(\cT_\bX)$ in (\ref{jiben}):
 \begin{equation*}
   G(\bX)  =  G(\cT_\bX) = PV(\bX) +  U(\bX) - TS(\bX) 
   = \sum_{i=1}^H \mu_i N_i(\bX) + \mu_eN_e(\bX).
 \label{gibbs}
\end{equation*}

\subsection{Converting Formula (\ref{jiben}) to Geometric Form 
(\ref{geo})}

Since every water molecule in $\cR_{\bX\, i}$ has contact with the surface $M_{\bX \, i}$,
   $N_i(\bX)$ is proportional to the area $A(M_{\bX\, i})$.   Therefore, there are
   $\nu_i > 0$, such that 
 \begin{equation}
   \nu_i A(M_{\bX\, i}) = N_i(\bX), \quad 1\le i \le H.  
 \label{area} 
 \end{equation}
   Similarly,  there will be an $a > 0$ such that $aV(\cT_\bX) = N_e(\bX)$.
   
 By the definition of $\cT_\bX$ and $\Omega_\bX$, we have roughly
 $V(\cT_\bx \! \setminus \! \Omega_\bX) = d_w A(M_\bX)$.   Thus
  \begin{equation}
 N_e(\bX) = aV(\cT_\bX) = a[V(\Omega_\bX) + V(\cT_\bX \!\setminus \! \Omega_\bX)]
 = a V(\Omega_\bX) + ad_w A(M_\bX).
  \label{tiji}
  \end{equation}
  Substitute (\ref{area}) and (\ref{tiji}) into (\ref{jiben}), we get  (\ref{geo}).
  
  We are applying fundamental physical laws directly to protein folding.  The question is, can we
do so?   We will try to check how rigorous is the derivation and ask that 
are there any fundamental errors? We will also discuss possible
ways to modify the formula or the derivation.

\subsection{How Rigorous Is The Derivation?}
   We adopted two common tools in physics, the first step of the Born-Oppenheimer
approximation in quantum mechanics and the grand canonic ensemble in statistical physics
to obtain formula (\ref{jiben}). 

\subsubsection{The Born-Oppenheimer Approximation}
 The Born-Oppenheimer approximation 
  ``treats the electrons as if they are moving in the field of fixed nuclei.
This is a good approximation because, loosely speaking, electrons move much
faster than nuclei and will almost instantly adjust themselves to a change in
nuclear position." \cite{popelier2000}.  Since the mass of a water molecule is
much less than the mass of a protein, we can extend this
approximation to the case of when $\bX$ changes the other articles,
electrons and water molecules, will quickly
adjust themselves to the change as well.

\subsubsection{The Statistic Physics in General and the Grand Canonic Ensemble in Particular}
  ``Up to now there is no evidence to show that 
 statistical physics itself is responsible for any mistakes," \cite[Preface]{dai2007}.
 Via the ensemble theory of statistical mechanics we consider only one protein molecule and particles in its
 immediate  environment, it is justified
 since as pointed out in \cite[page 10]{dai2007} ``When the duration of measurement is short, or the
 number of particles is not large enough, the concept of ensemble theory is still valid."
  And among different ensembles, ``Generally speaking, the grand canonic ensemble, with the least 
 restrictions, is the most convenient in the mathematical treatment."  \cite[page 16]{dai2007}.
In fact, we have tried the canonic ensemble and ended with a result that we have to really calculate
the eigenvalues of the quantum mechanics system.   

Our derivations only put together the two very common and sound practices: the Born-Oppenheimer approximation (only the
first step) and the grand canonic ensemble, and apply them to the protein folding problem.  As long as
protein folding obeys the fundamental physical laws,
there should not be any serious error with the derivation.

\subsection{Equilibrium and Quasi-Equilibrium}   A protein's structure will never be in
equilibrium, in fact, even the native structure is only a snapshot of the constant vibration
state of the structure.   The best description of conformation $\bX$ is given in 
\cite[Chapter 3]{bader1990}, we can simply think that a conformation $\bX$ acturally
is any point $\bY$ contained in a union of tiny balls centered at $\bx_i$, $i = 1, \cdots, M$.
In this sense, we can only anticipate a quasi-equilibrium description
(such as the heat engine, \cite[page 94]{bailyn}) of the thermodynamic states of the protein folding.  
This has been built-in in the Thermodynamical Principle
of Protein Folding.  So the quantities such as $S(\bX)$, $\Omega(\bX)$, and $G(\bX)$ can only 
be understood in this sense.  That is,
 observing a concrete folding process one will see a series conformations $\bX_i$, $i = 1, 2, 3, \cdots $.
  The Thermodynamic Principle then says that if we measure the
  Gibbes free energy $G(\bX_i)$ then eventually $G(\bX_i)$ will converge to a minimum value
  and the $\bX_i$ will eventually approach to the native structure.  While all the time, no conformation $\bX_i$
  and thermodynamic system $\cT_{\bX_i}$ are really in equilibrium state.

\section{Kinetics Formulae}
Let $\bx_i = (x_i, y_i, z_i)$, we can write $\bF = \grad_{\bx_i}G(\bX) = (G_{x_i}, G_{y_i}, G_{z_i})(\bX)$.
The calculation of  $G_{x_i}(\bX)$, for example, is via Lie vector field induced by moving the atomic 
position $\bx_i$.  In fact, any infinitesimal change of structure $\bX$ will induce a Lie vector field $\vec L: \bX
 \to \bbR^3$.  For example,  moving $\bx_i$ from $\bx_i$ to $\bx_i + (\Delta x_i, 0, 0)$
while keep other nuclear center
fixed will induce $L_{x_i} : \bX \to \bbR^3$, such that $\vec L_{x_i}(\bx_i) = (1, 0, 0)$ 
and $\vec L_{x_i}(\bx_j) = (0, 0, 0)$ for $j \ne i$.   Similarly we can describe $\vec L_{y_i}$ and
$\vec L_{z_i}$.   Then write $G_{x_i} = G_{\vec L_{x_i}}$, etc. and
\begin{equation}
  \grad _{\bx_i}G(\bX) = (G_{\vec L_{x_i}}, G_{\vec L_{y_i}}, G_{\vec L_{z_i}})(\bX),
\end{equation}

    Rotating around a covalent bond
$b_{ij}$ also induce a Lie vector field $L_{b_{ij}}: \bX \to \bbR^3$.  In fact if
  $\ba_i\ba_j$ form the covalent bond $b_{ij}$, then the bond axis is
  \begin{equation*}
  \bb_{ij} = \frac{\bx_j - \bx_i}{|\bx_j - \bx_i|}.   
  \end{equation*}
  If $\bb_{ij}$ is rotatable, i.e., 1. it is chemically allowed
  to rotate; 2. cutting off $\bb_{ij}$ from the molecular graph of $\bX$
  (see, for example, \cite[page 32]{bader1990})
  with two components, denoted all nuclear centers in one component by $R_{b_{ij}}$ and 
  others in $F_{b_{ij}}$.  We can rotate all centers in $R_{b_{ij}}$ around $\bb_{ij}$ for 
  certain angle while keep all centers in $F_{b_{ij}}$ fixed.   The induced Lie vector field $\vec L_{b_{ij}}$
  will be
  \begin{eqnarray}
    \vec L_{b_{ij}}(\bx_k) & = & (\bx_k - \bx_i)\wedge \bb_{ij}, \text{  if }\bx_k\in R_{b_{ij}}; \\
    \vec L_{b_{ij}}(\bx_k) & = & \vec 0, \text{ if } \bx_k \in F_{b_{ij}}.
   \end{eqnarray}
 Any such a Lie vector field $\vec L$ will generate a family of conformations $\bX_t=
 (\bx_{1\,t}, \cdots, x_{i\,t}, \cdots, x_{M\, t})$, where $\bx_{k \, t} = \bx_k + t \vec L(\bx_k)$,
 $k = 1, \cdots, M$.
 
The derivative $G_{\vec L}(\bX)$ is given by
\begin{equation}
  G_{\vec L}(\bX) = a\mu_e V_{\vec L}(\Omega_\bX) + ad_w\mu_e A_{\vec L}(M_\bX) 
  + \sum_{i=1}^H\nu_i\mu_i A_{\vec L}(M_{\bX \, i}),
\end{equation}
where
\begin{equation}
V_{\vec L}(\Omega_\bX) = - \int_{M_\bX} {\vec L}\but \vec N \ud \cH^2, \quad
A_{\vec L}(M_\bX) = - 2\int_{M_\bX} H ({\vec L}\but \vec N) \ud \cH^2,
\end{equation}
where $\vec N$ is the outer unit normal of $M_\bX$, $H$ the mean curvature of $M_\bX$, and
$\cH^2$ the Hausdorff measure.
Define $f_{t, i} : \bbR^3 \to \bbR$ as $f_{t\, i} (\bx) = \dist (\bx, M_{\bX_t \, i}) - \dist (\bx, M_{\bX_t}\! \setminus
\! M_{\bX_t\, i})$, and denote
\begin{equation}
  \grad_{M_\bX} f_{0, i} = \grad f_{0, i} - (\grad f_{0, i}\but \vec N)\vec N, \quad
   f'_{0, i} = \left . \frac{\bdy f_{t \,i}}{\bdy t}\right |_{t=0}, \quad
   \frac{\ud f_{0, i}}{\ud t} = \vec L \but \grad f_{0, i} + f'_{0, i},
 \end{equation}
 then let $\vec \eta $ be the unit outward conormal vector of $\bdy M_{\bX\, i}$ 
 (normal to $\bdy M_{\bX\, i}$ but
 tangent to $M_\bX$),
 \begin{equation}
  A_{\vec L}(M_{\bX\, i}) = - 2 \int_{M_{\bX\, i}} H (\vec L \but \vec N)\ud \cH^2
  + \int_{\bdy M_{\bX\, i}} \left [\vec L \but \vec \eta - \frac{\frac{\ud f_{0, i}}{\ud t}}
  {|\grad_{M_\bX}f_{0, i}|}\right ] \ud \cH^1.
 \end{equation}

The $\bX_t$ is all the
information we need in calculating the molecular surface $M_{\bX_t}$ \cite{connolly1983}.
   But the kinetic formula $G_{\vec L}(\bX)$ can help us quickly achieve a new conformation
   $\bY$ from $\bX$ without really calculating $\bX_t$.   For example, 
we can list all rotatable covalent bonds of the protein as $(\bb_1, \cdots, \bb_i, \cdots, \bb_L)$
and then simultaneously rotate them to get new conformations very quickly by moving along
the negative of the gradient
\begin{equation}
  (G_{\vec L_{\bb_1}}, \cdots, G_{\vec L_{\bb_i}}, \cdots, G_{\vec L_{\bb_L}})(\bX).
  \end{equation}
  To calculate the above formulae we actually have to translate them into formulae
  on the molecular surface $M_\bX$.  These translations are given in \cite{fj2008}, they are 
  calculable (all integrals are integrable, i.e., can be expressed by analytic formulae with variables
  $\bX$) and
  were calculated piecewisely on $M_\bX$.  If the rotation around $\bb_i$ with
  rotating angle $-sG_{\vec L_{b_i}}(\bX)$ be denoted as $\bR_i$, we can
  then get the new conformation $\bY_s = \bR_L \circ \bR_{L-1} \circ \ldots \circ \bR_1(\bX)$,
  where $s > 0$ is a suitable step length.  
   The order of rotations in fact is irrelevant, i.e.,
  by any order we will always get the same conformation $\bY_s$, as proved in
  \cite{fj2008}.

   This actually is the Newton's fastest desciending method, it reduces the Gibbs free energy $G(\bX)$
   most efficiently.  Afore mentioned simulations in \cite{fj2010} used this method.

\bibliographystyle{plain}

\bibliography{fang}

\end{document}